# Fast Hybrid PSO and Tabu Search Approach for Optimization of a Fuzzy Controller

**Nesrine TALBI[1,2] and Khaled BELARBI[2]**

**[1] Electronic Department, Jijel University
Jijel, ALGERIA**

**[2] Electronic Department, Mentoury University
Constantine, ALGERIA**

## Abstract

In this paper, a fuzzy controller type Takagi_Sugeno zero order is optimized by the method of hybrid Particle Swarm Optimization (PSO) and Tabu Search (TS). The algorithm automatically adjusts the membership functions of fuzzy controller inputs and the conclusions of fuzzy rules. At each iteration of PSO, we calculate the best solution and we seek the best neighbor by Tabu search, this operation minimizes the number of iterations and computation time while maintaining accuracy and minimum response time. We apply this algorithm to optimize a fuzzy controller for a simple inverted pendulum with three rules.

## 1. Introduction

Evolutionary algorithms, originally designed to optimize the parameters, have been shown to intervene as early in solving a problem. If in the case of optimizing parameters of a controller, the controller structure is given in advance, many studies show that evolutionary algorithms can be used to automatically obtain this structure. They can intervene at the design phase of the control system and optimize both the parameters and structure of the control system.

The particle swarm optimization (PSO) is a technique that evolutionary uses a "population" of solutions candidate to develop an optimal solution of the problem. The degree of optimality is measured by a fitness function defined by user ( Clerc et al. 2001, Dutot et al. 2002, Ji et al. 2007, Kennedy and al. 1995 and 2001). The PSO is different from other methods of evolutionary computation in order that members of the public called " particles " are scattered in space of the problem (Kennedy et al. 1995 and 2001).

The behavior of the swarm must be described from the point of view of a Particle (Kennedy et al. 2001, Omran 2004, Van den Bergh 2002, Venter et al. 2002).

From the algorithm, a swarm is randomly distributed in space research each particle also has a random speed.

PSO is easy to implement and there are few parameters to adjust, but it suffers from slow convergence in refined search stage (weak local search ability). In this context, we tried to solve the problem of convergence time with the combination of PSO and Tabu Search(TS) which is easy to implement and has the advantage of saving all the old solutions visited using the principle of memory to avoid backtracking (cyclical).

In this paper, we propose a hybrid PSO-TS algorithm to optimize a fuzzy controller for Takagi-Sugeno zero-order. The controller inputs are the error and derivative of error and the output is the control itself. The algorithm automatically adjusts the triangular membership functions of input and fuzzy singletons of output. This fuzzy controller is used for stabilization of a simple inverted pendulum.

This paper is organized as follows: Section 2 presents the general structure of fuzzy controller to optimize, In Section 3, the PSO is explained. Section 4 summarizes Tabou Search, and the hybrid algorithm of the hybrid Particle swarm optimization and Tabu Search is presented in section 5. Simulation results are presented in Section 6. Finally, Section 7 outlines our conclusions.

## 2. Fuzzy Logic Controller (FLC) to be optimized

A FLC is composed of a knowledge base, that includes the information given by the expert in the form of linguistic control rules, a fuzzification interface, which has the effect of transforming crisp data into fuzzy sets, an inference system, that uses them together with the knowledge base to make inference by means of a reasoning method and a defuzzification interface, that translates the fuzzy control action thus obtained to a real control action using a defuzzification method [1].

In this paper we will optimize a fuzzy controller Takagi-Sugeno zero-order. The controller inputs are the error $e(t)$ and the derivative of the error $\Delta e(t)$, and its output is the command $u(t)$.

## 2.1 Fuzzification

For flexibility in the implementation of the regulator, we must limit the universe of input and output at intervals determined by the normalization of input and output [-1,1], to do this, we use gains of adaptations to have the desired dynamic.

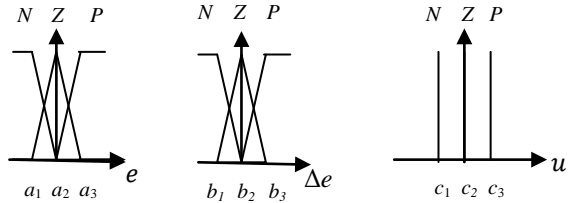

Fig. 1 Membership functions of inputs and output

The triangular membership functions are used for the Fuzzification of inputs (see Fig. 1). For example, the input $e(t)$ is fuzzied by following membership functions:

$$\mu_N(x) = \begin{cases} 1 & if\ x < a_1 \\ \dfrac{a_2 - x}{a_2 - a_1} & if\ a_1 \le x < a_2 \\ 0 & if\ x \ge a_2 \end{cases}$$

$$\mu_Z(x) = \begin{cases} \dfrac{x - a_1}{a_2 - a_1} & a_1 \le x < a_2 \\ \dfrac{a_3 - x}{a_3 - a_2} & a_2 \le x < a_3 \\ 0 & otherwise \end{cases} \qquad (1)$$

$$\mu_P(x) = \begin{cases} 0 & x < a_2 \\ \dfrac{x - a_2}{a_3 - a_2} & a_2 \le x < a_3 \\ 1 & x \ge a_3 \end{cases}$$

## 2.2 Fuzzy Inferences

The fuzzy rule base consists of three fuzzy rules according the Table 1. For the mechanism of inference, we used the method "min - max".

Table 1. Fuzzy Rule Base

| e \ de | N | Z | P |
|--------|---|---|---|
| N | N | - | - |
| Z | - | Z | - |
| P | - | - | P |

## 2.3 Defuzzification

we use the center of gravity for the defuzzification.

## 3. Particle Swarm Optimization (PSO)

PSO is a population-based stochastic optimization technique developed by Eberhart and Kennedy [2] that was inspired by social behavior of bird flocking or fish schooling. In PSO, each single solution is a "bird" in the search space. We call it "particle". All the particles have fitness values that are evaluated by the fitness function to be optimized, and have velocities that direct the flying of the particles. The particles fly through the problem space by following the current optimum particles [3].

In every iteration, each particle is updated by following two "best" values. The first one is the best solution and has achieved so far. This value is called *pbest*. Another "best" value that is tracked by the particle swarm optimizer is the best value, obtained so far by all particles in the population. This best value is a global best and called *gbest* [3].

During the iteration time *t*, the update of the velocity from the previous velocity to the new velocity is determined by Eq. (2). The new position is then determined by the sum of the previous position and the new velocity by Eq. (3):

$$V(t + 1) = wV(t) + c_1 * R1 * \big(pbest(t) - p(t)\big) \\ + c_2 * R2 * \big(gbest(t) - p(t)\big) \qquad (2)$$

$$p(t + 1) = p(t) + V(t + 1) \qquad (3)$$

$V$ is the particle velocity, The variable $w$ is called as the inertia factor, $p$ is the current solution, and $pbest$ and $gbest$ are defined as stated before. $R1$ and $R2$ are the random numbers uniformly distributed within the interval [0,1]. The variables $c_1$, $c_2$ are learning factors.

In order to guide the particles effectively in the search space, the maximum moving distance during one iteration is clamped in between the maximum velocity $[-Vmax, Vmax]$ given in Eq. (4) and similarly for its moving range given in Eq. (5) [4]:

$$V(t) = sign(V(t)) \min(|V(t)|, Vmax) \qquad (4)$$

$$p(t) = sign(p(t)) \min(|p(t)|, pmax) \qquad (5)$$

## 4. Tabu search (TS)

Tabu search is a higher level heuristic algorithm for solving combinatorial optimization problems. It is an iterative improvement procedure that starts from any initial solution and attempts to determine a better solution. TS was proposed in its present form a few years ago by Glover [5]. It has now become an established optimization approach that is rapidly spreading to many new fields. The flowchart of TS algorithm procedure is shown in Fig. 2 [6].

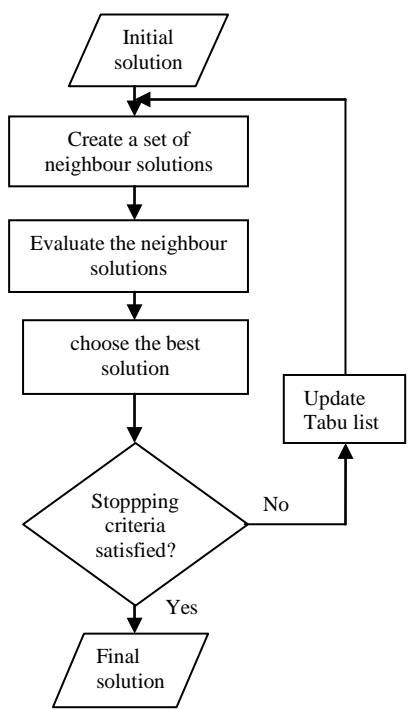

Fig. 2 Flowchart of a standard TS algorithm

## 5. Hybrid PSO-TS learning algorithm

To accelerate the convergence of PSO, it was proposed to combine the PSO algorithm with Tabu search to find a better solution in a minimum computation time and accuracy.

At each iteration of PSO, we calculate the best solution, then we will search its best neighbor by TS on minimizing a certain criterion (objective function) is the mean square error (MSE) calculated by the following equation:

$$MSE = 1/nT \sum_{k=1}^{n} e(k)^2 \qquad (6)$$

Where: $n$ is the total number of samples and $T$ the sampling time, $e(k) = r(k) - y(k)$ is the difference between the value of the desired output $r(k)$ and the value of the measured output $y(k)$ process under control.

Before turning to the hybrid algorithm PSO-TS, we must first introduce the particle on which the parameters are encoded to optimize. These are the parameters of each membership function of inputs, and conclusions of fuzzy rules.

According to Fig. 3 the particle will consist of nine parameters which are the modal values of membership functions of input and output fuzzy singletons, respecting the following constraint:

$$\begin{cases} a_1 < a_2 < a_3 \\ b_1 < b_2 < b_3 \\ c_1 < c_2 < c_3 \end{cases}$$

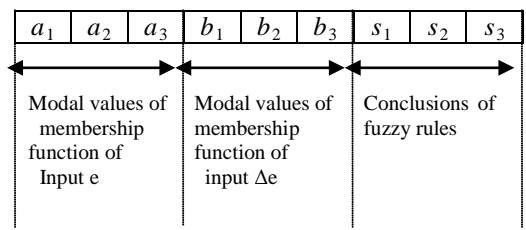

Fig. 3 *Particle* structure of PSO-TS

The pseudo-code of the procedure PSO-TS is as follows:

```
Begin PSO
    Initialize R1, R2, c₁, c₂, w
    Initialize Vmin, Vmax, pmin, pmax
    Initialize Particles randomly

    Do
        For each particle
            Calculate fitness value by Eq. (6)
                If the fitness value is better than the best
                fitness value (pbest) in history
            set current value as the new pbest
        End
          Choose the particle with the best fitness value of
          all the particles as the gbest
        For each particle
            Calculate particle velocity, V, according Eq. (2)
            Normalize V by Eq. (4)
```

   Update particle position, $p$, according Eq. (3)
   Normalize $p$ by Eq. (5)
End

 Begin Tabu Search
  Set the iteration counter $k = 0$
  Initialize the tabu list $T = 0$
  Initialize the aspiration function
  Do
   For each particle
    Generate solutions randomly
    Evaluate each neighbor by Eq. (6), and
    choose the best neighbor $p$
    Update the Tabu list and aspiration function
   End
  While the list tabu is not full or iteration is  not
  reached
 End

 While maximum iterations is not reached
End

## 6. Application

We will apply the hybrid PSO-TS algorithm to optimize a fuzzy controller Takagi-Sugeno zero-order. This controller will be used to control a simple inverted pendulum modeled by Eq (7).

$$\ddot{\theta} = \frac{-u\cos\theta + b\ \cos\theta\dot{x} - m\ l\ \dot{\theta}^2\sin\theta - (M+m)\ g\ \sin\theta}{l\left(\frac{4}{3(M+m)} - m\cos^2\theta\right)} \quad (7)$$

Where $\theta$ is the angular position and $u$ the control signal.

$g = 9.8\ m/s^2$ is the gravity constant, $m_c$ (mass of the cart) 1 kg, $m_p$ (mass of the pendulum) 0.1 kg, $l$ (length of the pendulum) 0.5 m.

The objective is to maintain the broom in the upright position by means of the force $f$ for any initial position of the broom without regard to the position and velocity of the cart.

The block diagram (see Fig. 4) shows the strategy optimization of the controller:

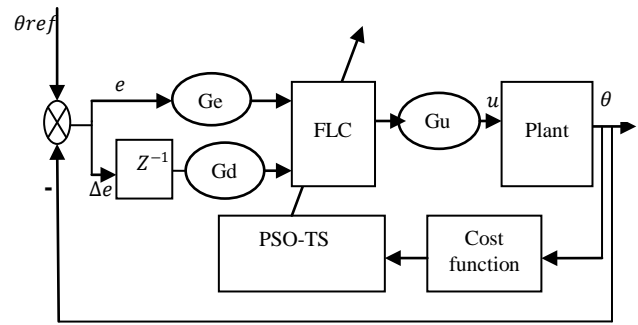

Fig. 4 Optimization of Fuzzy Controller

The cost function is calculated by Eq. (6) with a nominal model. The initial conditions are: $\theta(0) = 0.22\ rad, \dot{\theta}(0) = 0\ rad/sec$.

The PSO-TS algorithm was run for 2 generations. After each generation of PSO, TS was run for 5 iterations, The last generation contained a set of stable and performance satisfying solutions with controllers having three rules extracted for analysis.

Fig. 6 shows the form of membership functions of inputs and output after optimization and Fig. 7 shows the evolution of the cost function. The PSO-TS algorithm converges in a computation time equal to 2.88 sec with a response time of 0.2 sec < ts < 0.8 sec.

To demonstrate the robust ability of the selected fuzzy controller with proposed method method, different initial conditions (0.22 rad < $\theta$ < 0.8 rad) are shown in Fig. 5. The simulation result shows that the inverted pendulum is successfully controlled by the PSO-TS algorithm, and the angle of pole is quickly converged toward the balance.

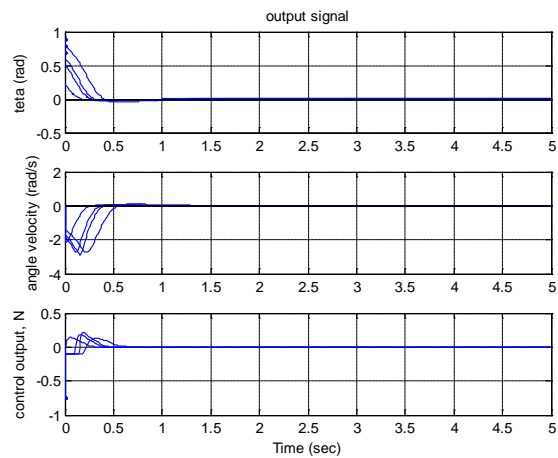

Fig. 5. the pendulum response to optimal fuzzy  controller for differents initial conditions.

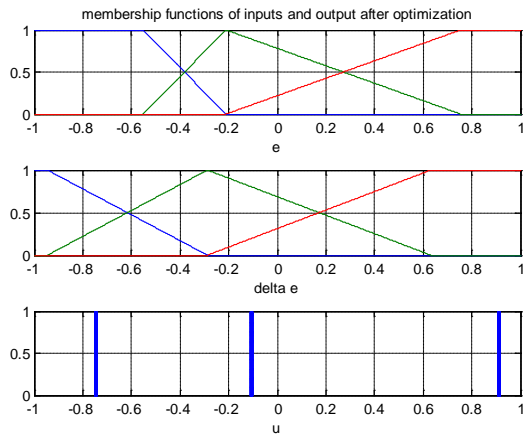

Fig. 6 Arrangement and forms of membership functions of the premises and conclusions after optimization

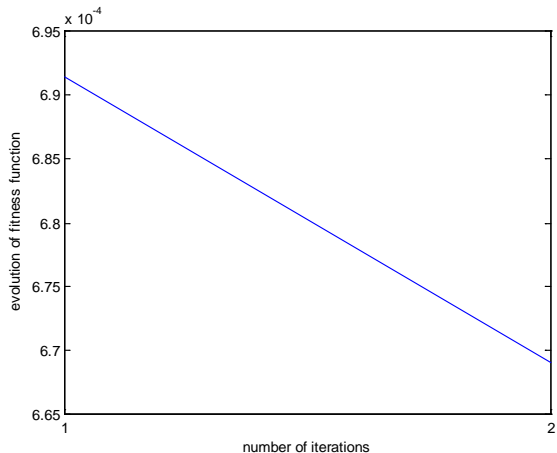

Fig. 7. Fitness function during iterations

## 7. Conclusion

In this paper, the PSO- TS hybrid algorithm is used to set the parameters of membership functions of inputs and conclusions of rules of fuzzy controller for Takagi-Sugeno zero-order, the latter is used to stabilize a inverted pendulum.

Simulation results have shown that the application of the proposed algorithm has improved the calculation time and the response time while optimizing the accuracy and simplifying the structure of this controller. We used only three fuzzy rules and a minimum number of generations for the implementation of this algorithm. The robustness test has proven the reliability of fuzzy controller optimized.

**Nesrine TALBI** was born in Jijel, ALGERIA in 1978, received her engineer and M.Sc degrees in Electronics, option: control systems from the Mentoury University, Constantine, Algeria. She is currently working toward the Ph.D. degree in control engineering at the University of Constantine. She is a Lecturer in Department of Electronic, Faculty of Sciences and Technology, Jijel University, ALGERIA.

**Khaled BELARBI** obtained his engineer degree polytechnic school, Algiers, Algeria, and MSc and PhD in control engineering both from Control System Center UMIST, Manchester, UK. He is currently a professor of electrical engineering with the Department of Electronics, Faculty of Engineering, University of Mentoury, Constantine, ALGERIA, and with the Laboratory of Automatic and Robotic at the same Department.His current interests are in predictive control, fuzzy control, intelligence control and neural control.